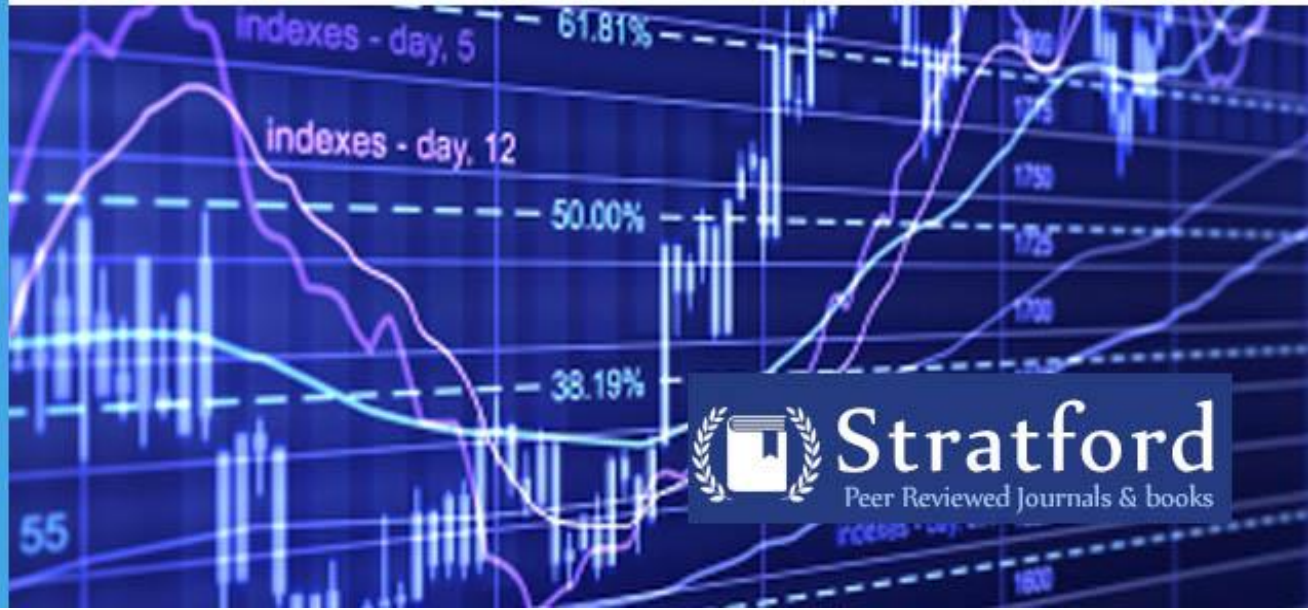

# Effect of Interest Payments on External Debt on Economic Growth in Kenya

Sammy Kemboi Chepkilot

ISSN: 2617-5800

# Effect of Interest Payments on External Debt on Economic Growth in Kenya

**Sammy Kemboi Chepkilot**

**Postgraduate, School of Economics, University of Nairobi**

Email: infokemboi@gmail.com

## Abstract

In Kenya, Interest on payment on external debt (% of exports of goods, services and primary income) has been on the increase since 2010 to 2015 while GDP growth experienced a slight decline from the year 2010 to 2015. There is concern among policymakers that the rapid increase in external debt in developing countries such as Kenya has the potential of eroding the country's sovereign rating, particularly if it is not supported by proportionate growth in the size of the economy. The purpose of this study was to investigate on the effect of interest payments on external debt on economic growth in Kenya. The study utilized secondary data for 25 years, that is, period 1991-2015 for GDP growth (annual %) and Interest payments on external debt (% of exports of goods, services and primary income). Data analysis was done using Eviews version 9 Software. A critical p value of 0.05 was used to determine whether the model was significant or not. The results on the analysis of the variance (ANOVA)-F statistics indicate that the model was statistically significant. This implies that interest payment on external debt is a good predictors of GDP growth. This was supported by an F statistic of 14.808 and the reported p value (0.0008) which was less than the conventional probability of 0.05 significance level. Regression of coefficient results shows GDP growth (annual %) and LOG Interest payment on external debt (% of exports of goods, services and primary income) are negatively and significantly related. Based on the findings above, the study concluded that when the country decide to service Interest payment as a result of external debt, might disturb the economic growth of the country. Based on the findings and conclusion above, the study recommends for government that future plans should ensure to take external borrowings which its rate is not higher than the interest rate payment. Also for policy implication, the government should not let interests on payment to accrue so much that it can compromise the growth of the economy.

## 1.0 Introduction

Interest payments are actual amounts of interest paid by the borrower in foreign currency, goods, or services in the year specified. This item includes interest paid on long-term debt, IMF charges, and interest paid on short-term debt (Boboye & Ojo, 2012). Long-term external debt is defined as debt that has an original or extended maturity of more than one year and that is owed to nonresidents by residents of an economy and repayable in foreign currency, goods, or services.

Short-term external debt is defined as debt that has an original maturity of one year or less (Nord, Harris, & Giugale, 2013).

Economists always sought out to find the ways through which a country can achieve long-lasting sustainable economic growth. The repayment of debt in the form of principal and interest payments which is cumulatively known as debt service payments is identified to be a serious threat to economic growth of any country, by keeping all other factors constant (Shabbir, 2012). Countries take debt from the external sources for many reasons i.e. their income is low, they are having budget deficit or they are having low investments in their country on conditions to repay them with certain obligations. This repayment or debt servicing creates problems for many countries especially for low income countries because a debt has to be serviced more than the actual amount it was taken for. Large debt service payments impose a number of constraints on a country's growth scenario. It drains out countries' limited resources and restricts financial resources for domestic need of development (Clements *et al*, 2003).

The 2008/2009 global financial crisis created concern about the risk of debt crisis in developing countries, with international finance institutions such as the World Bank and IMF stressing the importance of managing external debt obligations (World Bank, 2013). The global financial crisis and the ensuing economic recession created a debt crisis in developed countries, characterized by soaring public debt in the United States, sovereign debt crisis in Europe, and contagion effects in Asia, Africa, and Latin America (Miller & Foster, 2012). The debt ratio in the United States rose from 60 % to almost 100 % of GDP. In Japan, the debt ratio rose by 50% of GDP. In the United Kingdom, the debt ratio rose by 40% in 2007 and increased to 84% of GDP by the end of 2011 (Nautet & Meensel, 2012). In the euro area, the debt ratio is set to rise from 66% in 2007 to 88 % in 2011 (Nautet & Meensel, 2012), while Ireland recorded almost 90% of GDP.

In Sub-Saharan Africa, the 1950s and 1960s were described as the Golden Years', characterized by high and internally generated economic growth (Muhanji, 2010). External debt was comparatively small until the transition to debt-led growth began in the 1970s primarily as a result of the oil crisis (Boboye & Ojo, 2012). The 1973/74 oil price increase by Organization of Petroleum Exporting Countries (OPEC) led to general deterioration in the external payments position of the oil importing developing countries and forced many developing countries to borrow heavily, leading to an increase in the volume of international indebtedness and debt servicing liabilities (Boboye & Ojo, 2012).

In Kenya, despite improvement in revenue performance, data for the period 2009-2013 shows that on average, growth of expenditure at 18.7% outpaces revenue growth rate at 15.5% (IAS, 2014; KPMG, 2014). This deficit has increased reliance to external borrowing to fund the country's development activities (Stiglitz, 2001), hence the need to undertake cautionary austerity measures (IAS, 2014; KPMG, 2014). There are concerns that the rapid increase in external debt has the potential of eroding the country's sovereign rating, particularly if it is not supported by proportionate growth in the size of the economy (Nord, Harris, & Giugale, 2013).

Interest on payment on external debt (% of exports of goods, services and primary income) has been on the increase since 2010 to 2015 while GDP growth experienced a slight decline from the year 2010 to 2015 as shown in figure 1 below. There is concern among policymakers that the rapid increase in external debt in developing countries such as Kenya has the potential of eroding the country's sovereign rating, particularly if it is not supported by proportionate growth in the size of the economy (Nord, Harris, & Giugale, 2013).

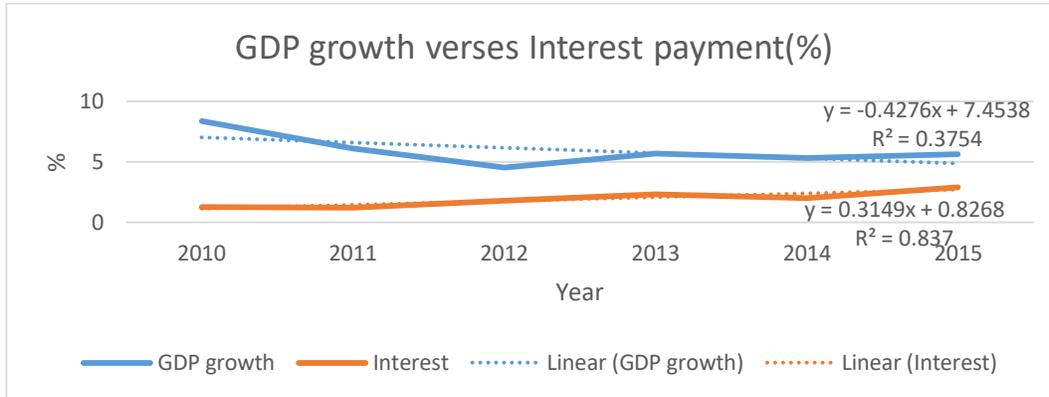

**Figure 1: Trends of GDP and Interest Payment for Kenya between the year 2010 and 2015**

### 1.1 Statement of the Problem

In Kenya, Interest on payment on external debt (% of exports of goods, services and primary income) has been on the increase since 2010 to 2015 while GDP growth experienced a slight decline from the year 2010 to 2015(WB,2015). There is concern among policymakers that the rapid increase in external debt in developing countries such as Kenya has the potential of eroding the country's sovereign rating, particularly if it is not supported by proportionate growth in the size of the economy (Nord, Harris, & Giugale, 2013). Kenya's public debt surged to 1.9 trillion according to the Quarterly Economic and Budgetary Review (October, 2013), with gross public debt increasing from Sh 1.633 trillion at the end of June 2012 to Sh 1.894 trillion by June 30, 2013, comprising of 44.5% external and 55.5% domestic debt. Overall debt levels have increased from 42.1 per cent of gross domestic product (GDP), or the value of the goods and services produced by the country in a year, in the 2012-13 financial year to 55.1 per cent of GDP in 2015-16. Total public debt now stands at around Sh3.2 trillion (WB report, 2016).Therefore this study sought to investigate on the effect of interest payments on external debt on economic growth in Kenya.

Most of local studies have been conducted on the effect of external debt on economic growth but scanty studies if they exist have focused on the effect of interest payments on external debt on economic growth in Kenya. Therefore this study was conducted so as to bridge the existing knowledge gap.

### 1.2 Research Objective

The general objective of this study was to establish the effect of interest payments on external debt on economic growth in Kenya.

### 1.3 Research Hypothesis

**H₀:** Interest payments on external debt does not have a significant effect on economic growth in Kenya.

## 2.0 Literature Review

### 2.1 Theoretical Review: Debt overhang theory

The theory that underpinned this study was debt overhang theory. According to Krugman (1988), the debt overhang theory shows that if there is some likelihood that in the future debt will be larger than the country's repayment ability, expected debt-service costs will discourage further domestic and foreign investment because the expected rate of return from the productive investment projects will be very low to support the economy as the significant portion of any subsequent economic progress will accrue to the creditor country (Krugman, 1988). This eventually will further reduce both domestic and foreign investments and hence downsize economic growth.

### 2.2 Empirical Review

Atique and Malik (2012) carried out a study to examine the determinants of economic growth in Pakistan and the impact of domestic and external debt on the economic growth separately over the period 1980-2010. Using the Ordinary Least Square (OLS) approach to Cointegration, Unit Root Testing, Serial Correlation Testing, test for checking Heteroskedasticity and CUSUM test of stability, the researchers demonstrated that there is an inverse relationship between the domestic debt and economic growth. The external debt and economic growth was also found to be inversely related. These relationships were found to be significant as well. The findings concluded that the amount of external debt slows down the rate of economic growth more than the amount of domestic debt.

Boboye & Ojo (2012) investigated the effect of debt burden on economic growth and development. A regression analysis of OLS was used to analyze secondary data from Central Bank of Nigeria (CBN), Economical and Financial review, Business times, Financial Standard and relevant publication from Nigeria on variable like National Income, Debt Service Payment, External Reserves, Interest rate among others. The findings showed that the external debt burden had an adverse effect on the nation income and per capital income of the nation. The study recommended that debt service obligation should not rise above foreign exchange earnings and that debts should be used for appropriate profitable investments where they can generate reasonable amount of money to fund debt repayment.

Sharif et al, (2009) analyzed the impact of foreign debt and foreign debt servicing on the savings and investment expenditure of Pakistan and suggested that foreign debt servicing has a negative impact on the constructive activities which can increase the economic growth of Pakistan.

Kasidi & Said (2013) examined the relationship between external debt and economic growth in Tanzania over the period 1990-2010. Data was collected from the Bank of Tanzania (BoT), President's Office of Finance, Ministry of Finance, World Bank, and International Monetary Fund publications. The findings demonstrated that there is a significant relationship between external debt and debt service on GDP growth. The total external debt stock had a positive effect of 0.36939 on GDP growth, while debt service payment had a negative effect of 28.517 on the GDP growth. Cointegration tests showed that there is no long run relationship between external debt and GDP

Putonoi & Mutuku (2012) concentrated on the effects of domestic debt owing to the shifting composition of public debt in favour of domestic debt in Kenya. The study used advanced econometric techniques and quarterly time series data from 2000 to 2010. The Jacque Bera (JB) and Augmented Dickey-Fuller (ADF) tests were used to investigate the properties of the macroeconomic time series in the aspect of normality and unit roots respectively, while the long run relationship between the variables was investigated using the Engel-Granger residual based and Johannes VAR based cointegration tests. There was evidence of cointegration hence an error correction model has been used to capture short run dynamics. The results demonstrated domestic debt expansion in Kenya, for the period of study. There was a positive and significant effect on economic growth. The study recommended that the Kenyan government should encourage sustainable domestic borrowing provided the funds are utilized in productive economic avenues

## 3.0 Methods

### 3.1 Data Type, Source, and Collection

The study utilized secondary data for 25 years, that is, period 1991-2015 for GDP growth (annual %) and Interest payments on external debt (% of exports of goods, services and primary income). (Attached in Appendix 1).

The data was obtained from World Bank. (*http://data.worldbank.org/data-catalog/world-development-indicators*). The primary World Bank collection of development indicators, compiled from officially-recognized international sources. It presents the most current and accurate global development data available, and includes national, regional and global estimates.

### 3.2 Model Specification

The main aim of this empirical enquiry is to determine whether interest payment on external debt affects economic growth in Kenya. According to Sala-i-martin (1997), economic theories are not sufficiently able to underpin the exact determinants of growth. A simple regression model was used to check on the relationship between interest payment on external debt (as % of exports of goods, services and primary income) and GDP growth. The natural logarithm of Interest payment was used so as to normalize the data.

$$GDPgr_t = \beta_0 + \beta_1 Ln(INTpayment_t) + \epsilon_t$$

Where,

GDPgr = Annual GDP growth

Ln(INTPayment) = Natural logarithm of Interest payment on external debt (as % of exports of goods, services and primary income).

In the model, $\beta_0$ = the constant term while the coefficient $\beta_1$ was used to measure the sensitivity of the GDP growth to unit change in the predictor variable, Interest payment on external debt (% of exports of goods, services and primary income). $\epsilon$ is the error term which captures the unexplained variations in the model.

Data analysis was done using Eviews version 9 Software. A critical p value of 0.05 was used to determine whether the model was significant or not.

### 3.3 Diagnostic Tests

### 3.3.1 Testing Normality

The study used the graphical method (Histogram) and Jarque-Bera test for normality to ensure that residuals of regression models are normally distributed. This is to ensure that the variables used in the analysis are distributed normally. It is considered to be normally distributed if the probability is greater than 0.05. If the data is not normally distributed, it is recommended that natural log of the variables is to be used. Results revealed that the data for GDP growth was normally distributed (jq=1.024, p=0.599) while for interest payment was not normally distributed (jq=11.583, p=0.003) but after using a **natural logarithm**, it became normally distributed (jq=1.584, p=0.452).as shown in Figure 2.

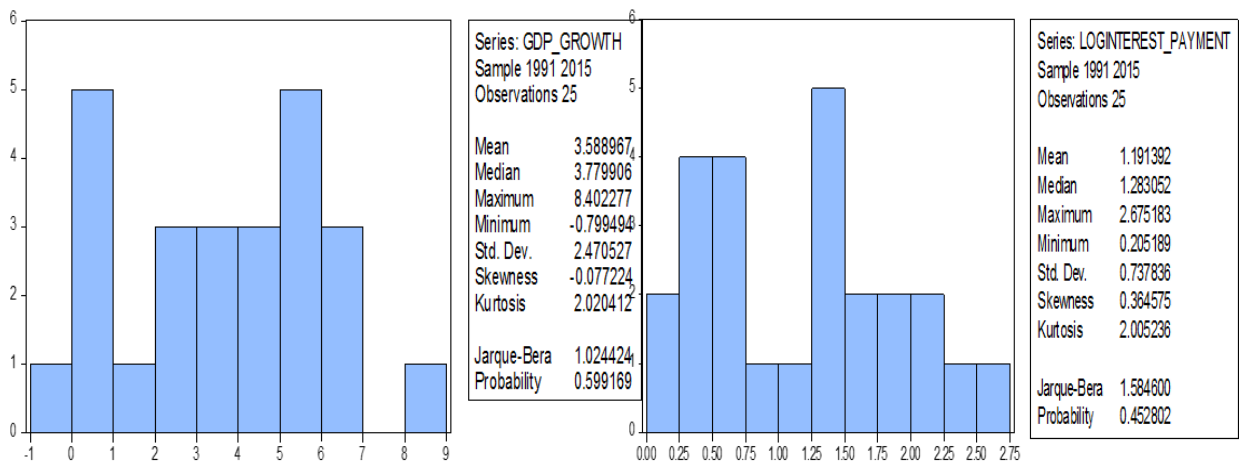

**Figure 2: Normality test using Jarque-Bera statistic**

### 3.3.2 Test for Autocorrelation:

The data was tested for autocorrelation using the Durbin- Watson Statistic method. The value for Durbin- Watson ranges from 0 to 4. The criterion indicates that a value of 2 shows that there is no autocorrelation. A value towards zero shows the presence of a positive correlation. A value towards four shows the presence of a negative autocorrelation. From the results of this study, the Durbin Watson value1.811 which was closer to 2 and thus there were no cause for alarm.

### Table 1: Durbin Watson Results

|  | **Statistic Value** |
|---|---|
| Mean dependent var | 3.588967 |
| S.D. dependent var | 2.470527 |
| Akaike info criterion | 4.268897 |
| Schwarz criterion | 4.366407 |
| Hannan-Quinn criter. | 4.295942 |
| **Durbin-Watson stat** | **1.811428** |

## 4.0 Results

### 4.1 Descriptive Statistics

Results in table 2 below indicate the descriptive statistics of GDP growth (annual %) and Interest payment on external debt (as a % of exports of goods, services and primary income). As indicated in the table, the mean of GDP growth for the period 1991 to 2015 was 3.5889 with a standard deviation of 2.470 indicating a small variability in GDP growth over time. It's Minimum and Maximum values were-0.7994 and 8.402 respectively. Further, results showed that the mean of Interest payment on external debt (as a % of exports of goods, services and primary income) for the period 1991 to 2015 was 4.317 with a standard deviation of 3.489 indicating a large variability in interest payment (%) over time. It's Minimum and Maximum values were 1.227 and 14.515 respectively.

**Table 2: Descriptive statistics results**

| Descriptive statistics | GDP_GROWTH (annual %) | INTEREST_PAYMENT (%) |
|---|---|---|
| Mean | 3.588967 | 4.317146 |
| Median | 3.779906 | 3.607635 |
| Maximum | 8.402277 | 14.51500 |
| Minimum | -0.799494 | 1.227757 |
| Std. Dev. | 2.470527 | 3.489507 |
| Skewness | -0.077224 | 1.470477 |
| Kurtosis | 2.020412 | 4.572056 |
| Jarque-Bera | 1.024424 | 11.58393 |
| Probability | 0.599169 | 0.003052 |
| Sum | 89.72416 | 107.9286 |
| Sum Sq. Dev. | 146.4841 | 292.2399 |
| Observations(years) | 25 | 25 |

### 4.2 Trend Analysis

Figure 2 below shows the trend analysis of GDP growth (annual %) and Interest payment on external debt (% of exports of goods, services and primary income) The trend analysis is conducted so as to help establish the movement of the variables under study and therefore help in determining their stationarity as the trend analysis graphically indicates the pattern of movement in the variables. Results shows that between the year 1991 and 2010, GDP growth has been on an increase while interest payment on external debt has been on a decline. But between the years 2010-2015 Interest on payment on external debt (% of exports of goods, services and primary income) increased while GDP growth experienced a slight decline from the year 2010 to 2015 as shown in figure 3.

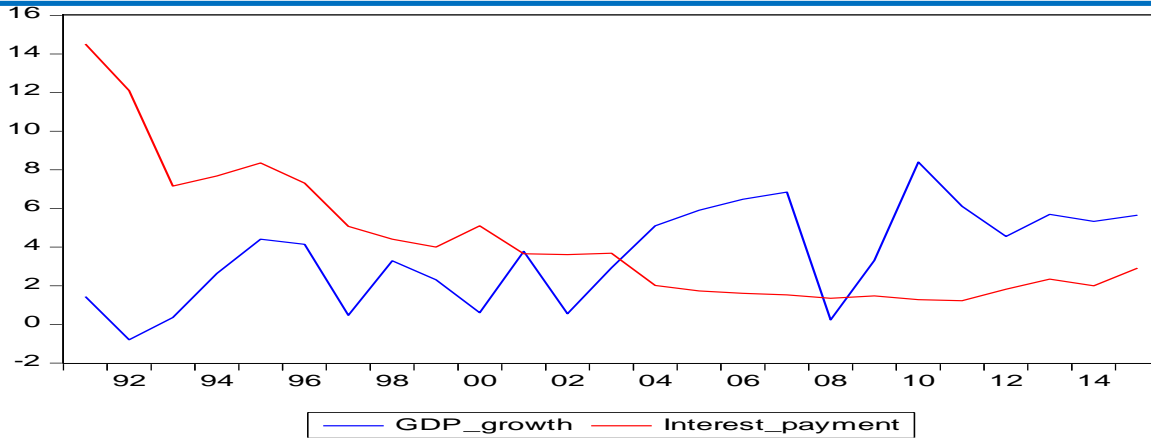

**Figure 3: Trends Analysis of GDP Growth (Annual %) and Interest Payment (%) between the year 1991 and 2015**

### 4.3 Correlation Analysis

Correlation analysis is the statistical tool that can be utilized to determine the level of association between two variables (Levin & Rubin, 1998). This analysis can be seen as the initial step in statistical modelling to determine the relationship between the dependent and independent variables. Prior to estimating the simple regression model, a correlation matrix was developed to analyze the strength of association between GDP growth (annual %) and interest payment (% of exports of goods, services and primary income).

Table 3 below presents the results of the correlation analysis. The results indicated that there was a negative and a significant association between GDP growth (annual %) and interest payment on external debt (% of exports of goods, services and primary income) (r=-0.6258, p=0.0008).

**Table 3: Overall correlation matrix results**

| Correlation t-Statistic Probability | GDP_GROWTH | Ln(INTEREST_PAYMENT) |
|---|---|---|
| GDP_GROWTH | 1.000000 | |
|  | ----- | |
|  | ----- | |
| Ln(INTEREST_PAYMENT) | -0.625827 | 1.000000 |
|  | -3.848088 | ----- |
|  | 0.0008 | ----- |

### 4.4 Regression Analysis

The R square of 39.16% shows that LOG interest payment on external debt (% of exports of goods, services and primary income) was found to be satisfactorily explaining the variation in GDP growth (%). The results on the analysis of the variance (ANOVA)-F statistics indicate that the model was statistically significant. This implies that interest payment on external debt is a good predictors of GDP growth. This was supported by an F statistic of 14.808 and the reported p value (0.0008) which was less than the conventional probability of 0.05 significance level. Regression of coefficient results in table 4 shows GDP growth (annual %) and LOG Interest payment on external debt (% of exports of goods, services and primary income) are negatively and significantly related (r=-2.0954, p=0.0008). This means that a 1 unit increase in interest payment (%) leads to a decrease in economic growth by 2.09 units.

Since the calculated p-value 0.0008< critical p-value, 0.05, the null hypothesis($H_0$) that Interest payments on external debt does not have a significant effect on economic growth in Kenya was rejected and thus arriving at the conclusion that Interest payments on external debt have a significant effect on economic growth in Kenya.

This finding is consistent with that of Kasidi & Said (2013) who examined the relationship between external debt and economic growth in Tanzania over the period 1990-2010 who found out that debt service payment had a negative effect on the GDP growth. Also seems to support the empirical assertion by Ruby (2012) on a study of the impact of external debt on economic of Bangladesh.

**Table 4: Regression Results**

| Variable | Coefficient | Std. Error | t-Statistic | Prob. |
|---|---|---|---|---|
| Ln(INTEREST_PAYMENT) | -2.095484 | 0.544552 | -3.848088 | 0.0008 |
| C | 6.085509 | 0.758872 | 8.019153 | 0.0000 |
| R-squared | 0.391660 | Mean dependent var | | 3.588967 |
| Adjusted R-squared | 0.365210 | S.D. dependent var | | 2.470527 |
| S.E. of regression | 1.968360 | Akaike info criterion | | 4.268897 |
| Sum squared resid | 89.11217 | Schwarz criterion | | 4.366407 |
| Log likelihood | -51.36121 | Hannan-Quinn criter. | | 4.295942 |
| F-statistic | 14.80778 | Durbin-Watson stat | | 1.811428 |
| Prob(F-statistic) | 0.000820 | | | |

$$GDPgrt = 6.0855 - 2.0954\, Ln(INTpayment)$$

Where,

GDPgr = Annual GDP growth

Ln(INTPayment) = Natural logarithm of Interest payment on external debt (% of exports of goods, services and primary income).

**5.0 Summary and Conclusion**

The relationship of GDP growth (annual %) and Interest payment on external debt (% of exports of goods, services and primary income) (1991 to 2015) in the graphical analysis shown there was some inverse relationship between GDP growth (annual %) and Interest payment on external debt (% of exports of goods, services and primary income).

The regression analysis results show that there is significant impact of Interest payment on external debt (% of exports of goods, services and primary income) on economic growth of Kenya. Further, results revealed that Interest payment on external debt (% of exports of goods, services and primary income) has a negative effect on economic growth of Kenya. This seems to support the empirical assertion by Kasidi & Said (2013) who examined the relationship between external debt and economic growth in Tanzania over the period 1990-2010 and found out that debt service payment had a negative effect on the GDP growth.

Based on the findings above, the study concluded that when the country decide to service Interest payment as a result of external debt, might disturb the economic growth of the country. This is evident from the regression results which showed that 1 unit increase in interest payment (%) leads to a decrease in economic growth by 2.09 units.

Based on the findings and conclusion above, the study recommends for government that future plans should ensure to take external borrowings which its rate is not higher than the interest rate payment. Also for policy implication, the government should not let interests on payment to accrue so much that it can compromise the growth of the economy.